\documentstyle[12pt,epsfig]{article}
\begin{document}
\textwidth 6.75in
\textheight 9in

\begin{center}
{\large\bf Study of
{$J/\psi\rightarrow\gamma \pi^{+}\pi^{-}\pi^{+}\pi^{-}$}}

\vskip 1.0cm
D.V.Bugg
\footnote{email address: david.bugg@stfc.ac.uk},   \\

\bigskip
Queen Mary, University of London, London E1 4NS, United Kingdom.

\end{center}

\vskip 0.5cm

\begin{abstract}
BES II data for the channel
$J/\Psi \to \gamma (\pi ^+\pi ^-\pi ^+\pi ^-)$
are analysed from a sample of 58M $J/\Psi$ hadronic interactions.
They reveal further detail in the $J^{PC} = 0^{-+}$ sector for
the $4\pi$ channel.
A broad $0^{-+} \to \rho \rho$ component extending from 1500 to 2400
MeV is fitted.
A narrower structure reveals a new $J^P = 0^-$ contribution
requiring a resonance of mass $1970 \pm 25(stat) \pm 60(syst)$
MeV and width $\Gamma = 210 \pm 25(stat) \pm 60(syst)$ MeV.
There is also tentative evidence for a $0^-$ resonance with mass
$1560 \pm (25) \pm 60(syst)$ MeV and width
$\sim 280$ MeV.
In the $0^{+}$ sector, there are clear contributions from $f_0(1500)$
and $f_0(1790)$.
The observed $f_0(1790)$ is visibly distinct from $f_0(1710)$ and
confirms the existence of $f_0(1790)$; its mass
$1800 \pm 25(stat) \pm 20(syst)$ MeV and width
$230 \pm 30(stat) \pm 25(syst)$ MeV are better determined here
than elsewhere.
 \end{abstract}

\section{Introduction}

\vskip 4mm
The analysis reported here was done in the years 2001--3 in
collaboration with three senior members of the BES collaboration.
The work was funded by the Royal Society and Queen Mary College
via a contractual agreement with the Chinese Academy of Sciences
and the BES collaboration.
This contract guaranteed access to BES 2 data for the purpose of
publications.
Since then, the BES management has refused to appoint internal
referees, hence blocking a collaboration publication.
The data illuminate the spectroscopy of $J^P=0^-$ mesons
and should be in the public domain.
There is a responsibility to publish work which has
been supported from public funds with approximately $\pounds 60,000$.
The wording of the present article is close to a document submitted
late in 2008 to the BES management, but some further
detail is added in Section 8.

There have been several earlier studies of
$J/\Psi \to \gamma (\pi ^+ \pi ^-\pi ^+\pi ^-)$.
The Mark III  \cite {MarkIII} and DM2 collaborations \cite {DM2}
observed a large $0^-$ signal and peaks at 1500, 1760 and 2100 MeV.
Later, Expt. E760 at Fermilab found three peaks in $\eta \eta$
\cite {E760} with masses and widths very similar to those in
$J/\Psi \to \gamma 4\pi$.
In the $\eta \eta$ channel, quantum numbers $J^P = 0^-$ are
forbidden by Bose statistics.
A re-analysis of the Mark III data \cite {Scott} led to the conclusion
that the peaks have quantum numbers $J^{PC} = 0^{++}$, but sit on a
large $0^{-+}$ signal extending over the whole mass range up to
2400 MeV.
BES I data led to similar conclusions \cite {BESI}, but added the
identification of a broad $2^{++}$ signal at $\sim 2$ GeV, decaying
to $f_2(1270)\sigma$ and $\rho \rho$.

Here statistics a factor 7 higher are presented from BES II, using
58M $J/\Psi$ hadronic interactions.
The new data reveal further detail in the $0^-$ sector.
There is clear evidence for a $0^-$ resonance at 1970 MeV.
This new feature alters details of what may be fitted in this
mass range to $f_0(2100)$ and the $J^P = 2^+$ amplitude.
There is also a strong $\rho \rho$ peak at $\sim 1600$ MeV.
Two alternative ways of fitting this peak are presented.
The first alternative is that it may simply be a threshold $\rho \rho$
cusp.
The second investigates evidence that the peak is actually resonant.
The distinction between the two alternatives depends rather delicately
on the phase of the $0^-$ amplitude.
In the $J^P = 0^+$ sector, there is clear evidence for
the existence of an $f_0(1790)$ which is definitely distinct from
$f_0(1710)$.

The layout of the paper is as follows.
Section 2 concerns technicalities of data selection and reduction of
experimental backgrounds.
There is a troublesome background at high $4\pi$ mass from
$\pi ^0\pi^+\pi^-\pi^+\pi^-$, mostly $a_2(1320)\rho$ and
$a_1(1260)\rho$.
This background has little consequence below 2 GeV, but does obstruct
separation of $J^P = 2^+$ and $0^+$ from 2 to 2.3 GeV.

Sections 3 and 4 present a fit to the $4\pi$ mass spectrum in slices
of mass.
The key point is that $J/\Psi$ radiative decays are pointlike,
because of the large $c\bar c$ mass and because photons interact at a
point.
The consequence is that only the lowest angular momentum states
contribute, making the selection of partial waves clean, see Fig. 3.
The dominant $J^P = 0^-$ channel, produced by an M1 transition,
varies in intensity as $P^3_\gamma$, where $P_\gamma$ is photon
momentum.
Superimposed on this broad signal is a narrow structure requiring
the phase variation of a resonance at $\sim 1970$ MeV.

For $J^P = 0^+$, there is a strong clear $f_0(1790)$ peak, definitely
distinct from $f_0(1710)$, see Figs. 3(b) and (c).
The $2^+$ channel is fairly weak.
There is a definite signal in the mass range 2--2.3 GeV, but it is
difficult to say whether it is a single broad component or may be
due to more than one resonance.
The experimental background is the limiting feature in this mass
range.
Section 5 shows detailed fits to $4\pi$ mass projections.

Section 6 concerns the question of how to fit the dominant very broad
$0^-$ component.
From other data, it is known that there are threshold peaks, see
Fig. 10, in $\omega \omega$, $\bar KK^*$, $\bar K^*K^*$ and
$\phi \phi$, as well as in the strong $\rho \rho$ signal observed here.
One or more of these peaks could be resonant, but it seems unlikely
there is one resonance per channel.
The alternative is that they are threshold cusps.
The formula fitting the broad $0^-$ component is based on this
conservative approach.
Section 7 concerns fitted branching fractions.

The fit to the $4\pi$ mass spectrum from 1500 to 1650 MeV is not
perfect in this approach.
This mass range can be fitted significantly better if the strong peak
observed in $\rho \rho$ (and elsewhere in $\eta \pi \pi$) is
resonant.
However, in the absence of strong interferences with other partial
waves, this evidence for a resonance is tentative, and is discussed
in Section 8.

There is one feature of the present data which is in strong
disagreement to an earlier publication claiming an $\eta (1760)$
resonance which completely dominates data on
$J/\Psi \to \gamma \omega \omega$.
It is suggested in Section 9 that an alternative way of fitting those
data is in terms of a simple cusp due to the $\omega \omega$
P-wave threshold.
Section 10 summarises conclusions and the Appendix outlines the SU2
relation concerning Section 9.

\section {Event Selection and Backgrounds}
The data presented here were taken with the BES\,II detector.
Full details of the detector and its upgrade are reported by
Bai et al. \cite {DetectA}, \cite {DetectB}.
It has cylindrical symmetry around the intersecting $e^+e^-$ beams.
Its essential features are
(i) a Main Drift Chamber (MDC) for the measurement of charged particles,
(ii) time of flight detectors with a $\sigma$ of 180 ps, and
(iii) a 12 radiation length Barrel Shower Counter comprised of
gas proportional tubes interleaved with lead sheets.
The MDC measures $dE/dx$.
Together with the time of flight detectors, it separates $\pi$ and $K$
up to $\sim 700$ MeV/c.
Outside the $\gamma$ detectors is a magnet providing a field of
0.4T.
This magnet is instrumented with muon detectors, but for  present
work they serve simply to reject $\mu ^+\mu ^-$ pairs.
The Main Drift Chamber provides full coverage of charged particles for
lab angles $\theta$ with $|\cos \theta | < 0.84$.

Events are selected initially by demanding four charged tracks in
the Main Drift Chamber with balancing charges, plus at least one photon
in the lead-scintillator detectors.
Any charged track with either time-of-flight or $dE/dx$ favouring kaon
identification is rejected.
Charged tracks are required to have a transverse momentum $>60$ MeV/c;
this reduces background from $J/\Psi \to \pi ^0\pi ^+\pi ^-\pi ^+\pi
^-$.
A geometrical cut is applied to ensure that all particles lie
inside a region well inside the known detector geometry, so as
to eliminate edge effects in detector performance and acceptance.
The vertex is required to lie within a cylindrical region of radius
2 cm and $\pm 20$ cm in length, with
respect to the known intersection point of the $e^+e^-$ beams.

Events have then been fitted kinematically to
$\gamma (\pi ^+\pi ^-\pi ^+ \pi ^- )$, using all photon showers in turn;
some photons may originate from interactions of charged particles in the
detector, but any photons within a cone of $6^\circ$ around a charged
particle are rejected.
Events with a $\chi ^2$ probability $>5\%$ are selected.
It is also required that $\chi ^2$ is better than that for likely
background channels:
(a) $4\pi$ ignoring photons,
(b) $2\gamma 4\pi$ if there is more than one photon and
$M(2\gamma)$ is within 50 MeV of the $\pi ^0$ mass.
A small detail is that background from $J/\Psi \to \omega \pi ^+\pi ^-$
has been rejected by a cut $|M_{\pi ^+\pi ^-\pi ^0} - M_\omega| > 25$
MeV, where the $\pi ^0$ is associated with missing momentum.
Also, to remove a small background from $J/\Psi \to \gamma K^0_S K^0_S$,
a double cut is made on the two $\pi ^+\pi ^-$ masses:
$|M_{\pi ^+\pi ^-}| > 25$ MeV.

The Monte Carlo generates 300K events to simulate the BES detector
and its acceptance; it uses the standard SIMBES package released in
February 2002, in its version as of March 2003.
During discussions with the BES management,
they have claimed that alterations to the calibration of the
time-of-flight system after March 2003 could affect the selection of
events.
I have investigated this claim thoroughly and it has no
foundation.
Firstly, there is excellent agreement between $dE/dx$ and
time-of-flight measurements.
Secondly, a search has been made for contamination of data due to
misidentified $\gamma K^+K^-\pi ^+\pi ^-$ events.
These are rather well separated from $\gamma 4\pi$ events by
kinematics as well as $dE/dx$ and time-of-flight.
For $60\%$ of events, the kinematic fit alone fails to fit
$\gamma K^+K^-\pi ^+ \pi ^-$ by a difference of $\chi ^2 >50$.
This is because production angles are sensitive to the different
masses of kaons and pions.

\begin{figure}
\begin{center}
\vskip -1cm
\epsfig{file=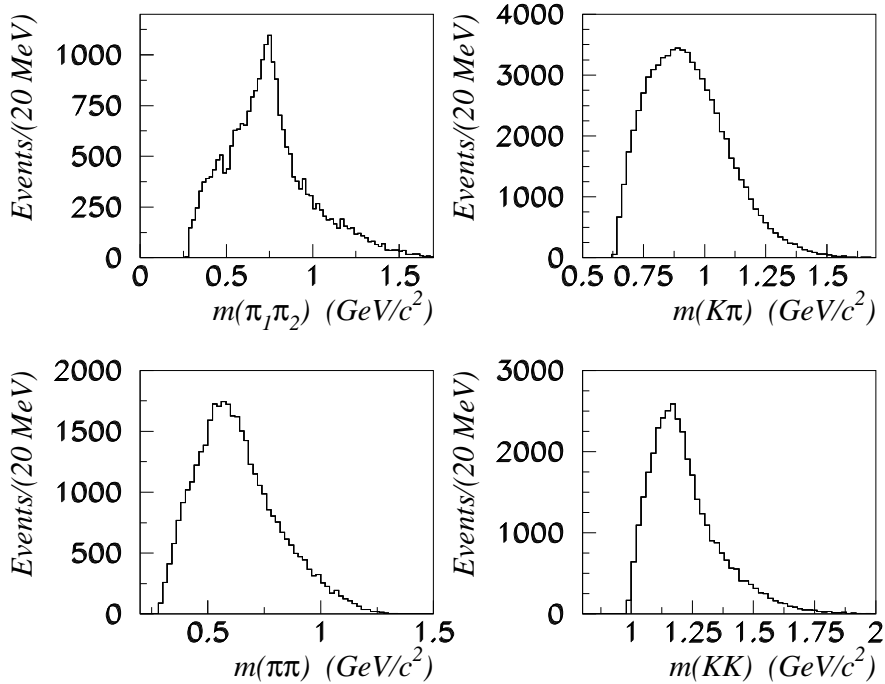,width=12cm}
\vskip -6mm
\caption[]{Mass projections for (a) $\pi \pi$ pairs from
events fitted to $\gamma \pi ^+\pi ^-\pi ^+\pi ^-$, taking $P(\pi ^+_1)
> P(\pi ^+_3)$ and $P(\pi ^- _2) > P(\pi_4^+)$; other panels refer to
events fitted to all combinations of $\gamma K^+K^-\pi ^+\pi ^-$: (b)
both $K\pi$ pairs, (c) $\pi ^+\pi^-$ pairs, (d) $K\bar K$ pairs.}
\end{center}
\end{figure}
Fig. 1(a) shows the mass distribution for $\pi _1^+\pi _2^-$ pairs
fitted to $\gamma 4\pi $ with momenta $P_1 > P_3$ and $P_2 >P_4$.
(Results for other $\pi ^+\pi ^-$ combinations are similar).
There is an obvious $\rho (770)$ signal; a small dip at 500 MeV
arises from the cut against $K^0_S$.
Next, these events have been fitted to all four combinations of
$\gamma K^+K^-\pi ^+\pi ^-$ after $dE/dx$ and time-of-flight cuts.
Events fitting with a $\chi ^2 < 50$  have the $K\pi$ mass
distribution shown in Fig. 1(b).
It is known from BES 1 data \cite {KKpp} that the
$\gamma K^+K^-\pi ^+ \pi ^-$ channel contains a very strong $K^*(890)$
signal.
There is no evidence for this in Fig. 1(b),
demonstrating that any contamination from $\gamma K^+K^-\pi ^+\pi ^-$
is $<1\%$ of good $\gamma 4\pi$ events.
Likewise there is no structure in the $K\bar K$ mass spectrum of
Fig. 1(d).
Fig. 1(c) shows the $\pi \pi$ mass spectrum from events surviving the
$dE/dx$ and time-of-flight cuts but fitted as
$\gamma K^+K^-\pi ^+ \pi ^-$.
One sees that the effect of the kinematic fit to the wrong
$\gamma K^+K^-\pi ^+\pi ^-$
hypothesis is to shift the $\rho (770)$ peak down in mass.

Other background processes and the acceptance for data are studied
using the standard SIMBES Monte Carlo simulation of the detector.
The main background arises from the $\pi ^0 \pi ^+\pi ^-\pi ^+\pi ^-$
channel after the loss of one photon.
This is well known from all earlier publications.
The $\pi ^0\pi ^+\pi ^-\pi ^+ \pi ^-$ channel has been studied by DM2
\cite {Augustin}.
It contains a strong $a_2(1320)\rho$ signal, a less
prominent $a_1(1260)\rho$ signal and thirdly a diffuse signal over all
of phase space.
The Monte Carlo has been used to generate $a_2(1320)\rho$ and
$a_1(1260)\rho$  events with the acceptance of the BES 2 detector.
Background from these channels in $\gamma 4\pi$ data peak strongly at
high $4\pi$ mass, because of the loss of low energy photons
associated with low energy $\pi ^0$.
This background dominates the $4\pi$ mass range $> 2400$ MeV/c$^2$,
allowing a firm determination of its magnitude.

\begin{figure}
\begin{center}
\vskip -1.5cm
\epsfig{file=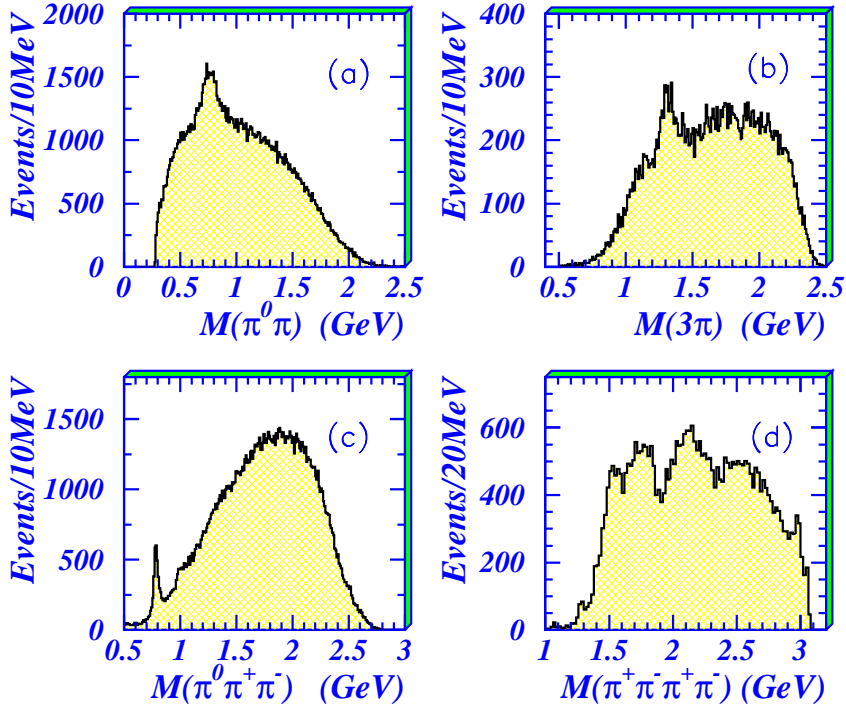,width=12cm}
\vskip -5mm
\caption[]{Mass projections: (a)--(c) refer to events where charged
particles alone are
consistent with the $\pi ^0 \pi ^+\pi ^-\pi ^+\pi ^-$ hypothesis.
(a) $M(\pi ^0 \pi ^{\pm })$,
(b) $M(3\pi )$ after selecting $M(\pi ^0 \pi ^{\pm})$
from 670 to 870 MeV, (c) $M(\pi ^0 \pi ^+\pi ^-)$, all events;
(d) $M(\pi ^+\pi ^- \pi ^+\pi ^-)$ after the full selection of
$\gamma \pi ^+\pi ^-\pi ^+\pi ^-$ events.}
\end{center}
\end{figure}

Both backgrounds produce clear $\rho (770) \to \pi ^0\pi ^{\pm}$,
see Fig. 2(a).
In these events, the background due to $a_2(1320)\rho$ leads to an
$a_2(1320)$ peak in the $\rho \pi$ mass distribution, shown in
Fig.~2(b).
Its magnitude is fitted freely and is consistent within errors with the
branching fraction quoted by the Particle Data Group (PDG) \cite {PDG}.
The $a_1$ is sufficiently broad that the $a_1(1260)\rho$ background
after cuts is consistent within errors with $5\pi$ phase space and
is fitted like that.
This background contribution is somewhat larger
than DM2 quote, suggesting that further diffuse background close to
$5\pi$ phase space is also present.
The combination of $a_1(1260)\rho$ and diffuse background is
again fitted freely to the $4\pi$ mass range above 2.4 GeV.
The sum of experimental backgrounds from $a_2(1320)\rho$ and
$a_1(1260)\rho$ is shown below by the dotted curve in Fig. 5(d).
Fig. 2(c) shows the observed $\pi ^0\pi ^+\pi ^-$ mass spectrum
in events fitted to $\pi ^0\pi ^+\pi ^-\pi ^+\pi ^-$.
The $\omega $ peak disappears after cuts used to select
$\gamma 4\pi$.

Fig.~2(d) shows the whole mass distribution for
$\pi ^+\pi ^- \pi ^+\pi ^-$ after the full event selection.
Peaks are visible below a mass of 2.4 GeV.
Above that, background dominates, so the analysis reported here is
confined to the mass range below 2400 MeV.
The total background up to 2.4 GeV is 27.3$\%$ of the whole
$\gamma 4\pi$ sample.

From Monte Carlo studies, estimated backgrounds from
$K^+K^-\pi ^+ \pi ^-$ and $\pi ^0K^+ K^-\pi ^+\pi ^-$ are $<2\%$;
the latter background is peaked at high $K^+K^-\pi ^+\pi ^-$
masses, like the $\pi ^0\pi ^+\pi ^-\pi ^+\pi ^-$ background,
and is mostly removed by selecting events with $4\pi$ mass
$< 2400 $ MeV.

\section {Introduction to formulae for partial waves}
Required channels are $\rho \rho$, $\sigma \sigma$ and
$f_2(1270)\sigma$.
There are several parametrisations of the $\sigma$
amplitude with increasing complexity as its couplings to $K\bar K$,
$\eta \eta $ and $4\pi$ have been unravelled \cite {Zou},
\cite {sigpole}, \cite {f01370}.
Only the coupling to $\pi \pi$ is important here.
All have been tried in the analysis, but there is little to choose
between them.
The first formula parametrises $\pi \pi$ data adequately and gives
slightly the best fit; it is the simplest and is used here.

There might be contributions from resonances decaying
to $a_2(1320)\pi$ and/or $a_1(1260)\pi$.
There is no evidence for $a_2\pi$ from this source in mass projections
of $\rho \pi$ after background subtraction.
Decays of $0^-$ to $a_1\pi$ are forbidden by parity conservation.
Adding decays of $f_0$ and $f_2$ to $a_1\pi$ from each resonance
one by one gives negligible improvements in log likelihood $(< 3)$.

The angular momentum in decays of resonances will be denoted by
$L$ and the angular momentum in the production process $J/\Psi \to
\gamma X$ will be denoted by $L'$.
The dominant process for production of $X$ with $J^P = 0^-$
requires $L'=1$.
The coupling of the photon to $c\bar c$ is a point interaction.
Because the charmed quarks are massive, the interaction of gluons with
$c\bar c$ is expected to be of very short range, with a
radius parameter of order $1/m_c \sim 0.1$ fm.
The intensity for $J/\Psi \to \gamma 0^-$ is then proportional to
$P^3$, (where $P$ is the momentum of the photon and hence the $4\pi$
system), multiplied by a very slowly varying form factor discussed
below.
The data are indeed close to the predicted $P^3$ dependence.
This has an important consequence: production with $L' > 1$ should be
strongly inhibited.
Higher $L'$ values were tried in the fit, and in all cases
are negligible.
This leads to a major simplification in the analysis.

For production of $J^P = 0^-$, the $L' = 1$ operator is represented
by the total momentum $K_X$ of $X$ in the $J/\Psi$ rest frame.
When it decays to $\rho _A \rho _B$, followed by
$\rho _A \to \pi _1 ^+\pi _2^-$ and $\rho _B \to \pi _3^+\pi _4^-$ the
vectors describing $\rho _A$ and $\rho _B$ are $K_A = k_1 - k_2$
and $K_B = k_3 - k_4$, where $k$ are momenta of pions in the rest
frame of each $\rho$.
[The 14 and 23 combinations also need to be added coherently
in the program].
The full formalism for Lorentz invariant tensors is presented in
Ref. \cite {formulae}.
It involves a Lorentz transformation of $K_A$ and $K_B$ to the
centre of mass of X.
Apart from this complication, the formula for production and
decay of $0^-$ is given by $K_A\wedge K_B.K_X$.
This form is highly distinctive, particularly the vital dependence
on $K_X$.

The states with $J^P=0^+$ and $2^+$ are produced with $L' = 0$.
The $0^+$ amplitude for $\sigma \sigma$ decays is isotropic;
for decays to $\rho \rho$ it is proportional to $K_A.K_B$.
The $2^+$ amplitude is given by the well-known spin 2 tensor:
$K^\alpha _A K^\beta _B -  (1/3)(K_A.K_B)$.
All three $J^P$ components are cleanly and simply identified.
There is no evidence for components with other $J^P$, e.g. $1^+$
and $2^-$.
A check on all amplitudes in the program is  that they are orthogonal
when integrated over $4\pi$ solid angle.

\begin{figure}
\begin{center}
\vskip -16mm
\epsfig{file=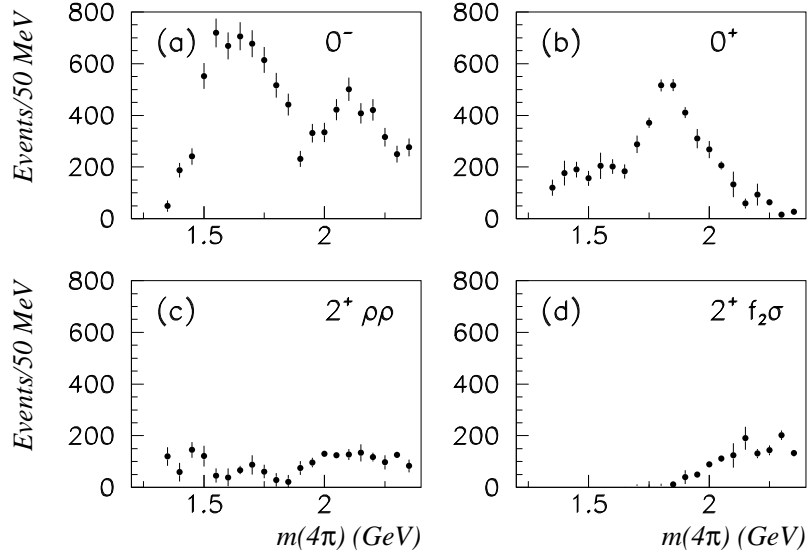,width=11cm}
\vskip -6mm
\caption[]{$J^P = 0^-$, $0^+$ and $2^+$ components from a fit in 50
MeV slices of $4\pi$ mass from 1350 to 2350 MeV. (b) shows
the sum of $\sigma \sigma$ and $\rho \rho$ signals;
the former dominates strongly.}
\end{center}
\end{figure}

\section {The Slice Analysis}
The data are fitted in 50 MeV bins of $4\pi$ mass and results
are shown in Fig. 3.
The $0^+$ amplitude peaks near 1790 MeV and has a shoulder near 1500
MeV.
For $J^P = 2^+$, decays to both $\rho \rho$ and $f_2(1270)\sigma$
are observed.
There is some cross-talk between these two final states, but both are
consistent with a broad peak at $\sim 2150$ MeV, plus a small
$f_2(1270) \to \rho \rho$ at masses 1270--1400 MeV.
The latter is weak and not well defined by present data.
Its magnitude is known from the analysis of $J/\Psi \to \gamma \pi ^+
\pi ^-$ and the $3.3\%$ branching fraction of
$f_2(1270) $ to $\rho ^0 \rho ^0$, and is fixed to this value in the
full fit described below; its phase is fitted freely.

From Fig. 3(a), the $J^P = 0^-$ amplitude appears to be made of two
components: peaks at 1600 and 2100 MeV.
However, the data cannot be fitted successfully like that.
The reason is that there is clear evidence for  phase variation
consistent with a resonance near 1970 MeV interfering with the
broad $0^-$ component.
A good fit is obtained with a peak at $\sim 1600$ MeV in the broad
component, dropping smoothly to 2400 MeV, plus an interfering component
over the mass range 1850 to 2200 MeV, as shown below in  Fig. 5(a).
The smooth drop in the broad component from 1600 to 2400 MeV is close
to the $P^3$ dependence expected for point-like production.

\begin{figure}
\begin{center}
\vskip -20mm
\epsfig{file=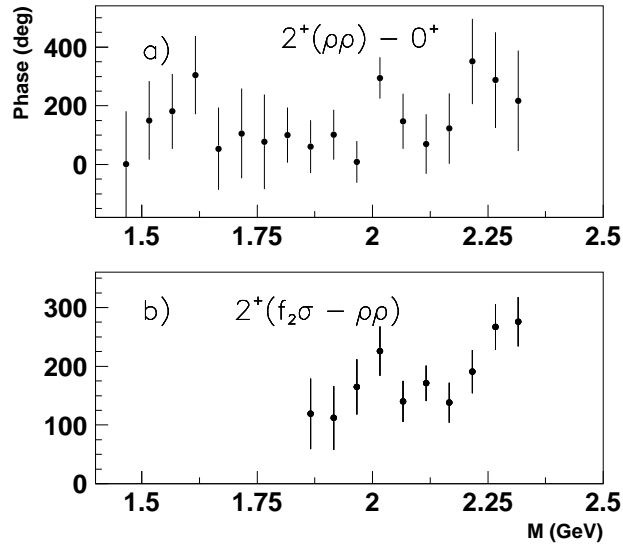,width=9cm}
\vskip -6mm
\caption[]{Differences in the slice fit between phases of
(a) $2^+ (\rho \rho )$ and $0^+$, (b)$ 2^+
[f_2(1270)\sigma ]$ and $2^+ (\rho \rho )$. }
\end{center}
\end{figure}

Although the magnitudes of the dominant components are well determined
in Fig. 3, their relative phases are poorly determined.
The final state with $J^P = 0^-$ is produced by an M1 transition
and that with $J^P = 0^+$ is produced by an E1 transition.
The program takes the trace over the two photon polarisations
allowed by Gauge Invariance and there is then no interference
between $0^-$ and $0^+$.
There is interference between $0^+$ and $2^+$ and between
$0^-$ and $2^+$, but it is fairly small in practice.
However, when the program takes the trace using all three
spins $0^-$, $0^+$ and $2^+$, there does remain some significant
information from details of angular correlations on the phase of the
$0^-$ amplitude and this will be discussed in detail in Section 8.

Fig. 4 displays information on phases from the slice fit after
integrating over all angles.
Errors for $0^-$, $0^+$ and $2^+$ are determined from the change in log
likelihood as the phase of each individual partial wave is moved in
steps.
Errors for the phase difference between $0^-$ and $0^+$ span most of the
range 0 to $360^\circ$, so these results are inconclusive and are not
shown.
The relative phase between $2^+$ and $0^+$ is shown in Fig.
4(a).
The two $2^+$ amplitudes are compatible within errors with a
constant or with a slowly rising phase difference in Fig. 4(b).
However, Fig. 4 is not able to display the delicate correlations
between all three spins if the angular information is used in addition.
This point will be discussed in Section 8.

\section {Full fit to data}
The dominant feature in the $4\pi$ mass spectrum is the broad
$J^{PC} = 0^{-+}$ signal in $\rho \rho$, extending
over the whole mass range from 1500 to 2400 MeV.
It will be described here as `the broad $0^-$'.
The rapid rise of the $0^-$ cross section at low mass is due
to the threshold for $\rho \rho$.

In Fig. 5(a), the upper histogram in each panel shows the overall fit
to the $4\pi$ mass projection.
The fit to the mass region 1500-1750 MeV is not perfect.
There appears to be one low point at 1610 MeV.
No available ingredient is narrow enough to fit a single bin.
The closest is $f_2(1565)$, but it has a full-width at half-maximum of
150 MeV.
Section 8 will present improvements if the 1600 MeV
$\rho \rho$  peak is fitted as a resonance rather than a threshold
cusp.

\begin{figure}
\begin{center}
\vskip -2cm
\epsfig{file=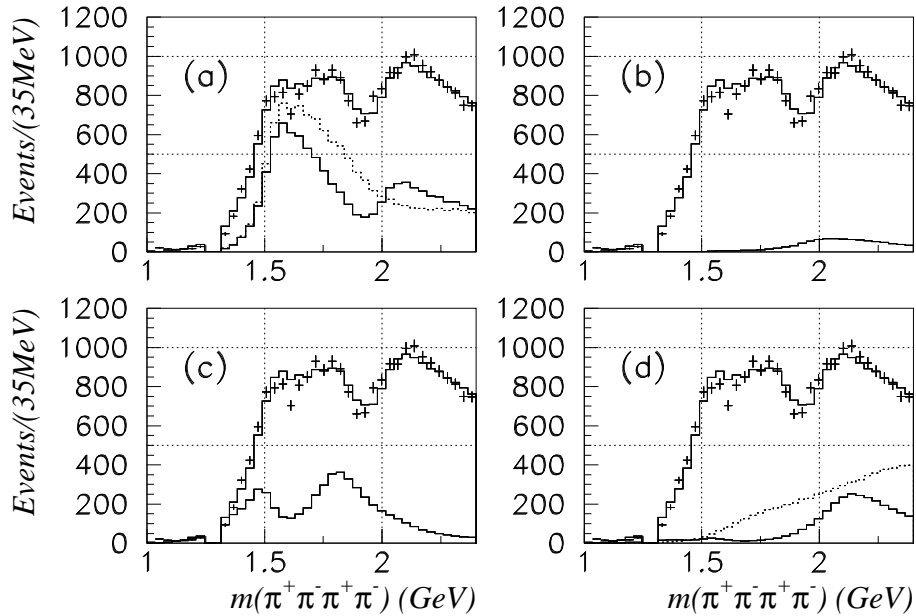,width=12.5cm}
\vskip -6mm
\caption[]{In each panel, the data points and upper histogram show
the full $4\pi$ signal for reference.
There is a cut from 1245 to 1325 MeV to remove $f_1(1285) \to 4\pi$.
Lower histograms show
individual components: (a) Full $0^-$ intensity (full histogram) and
from the broad $0^-$ alone (dotted); (b) $\eta (1970)$;
(c) coherent sum of all $0^+$; (d) coherent sum of all $2^+$ components
(full histogram) and background (dotted). }
\end{center}
\end{figure}

Fig. 5(a) shows as the lower full histogram the whole fitted $0^-$
component.
The dotted histogram shows the slowly varying $0^-$
component, which falls smoothly from 1560 to 2400 MeV.
Fig. 5(b) shows the component fitted as a $0^-$ resonance at
$1970 \pm 25(stat) \pm 60(syst)$ MeV
with $\Gamma = 210 \pm 25(stat) \pm 60(syst)$ MeV.
Systematic errors cover the whole range of fits and variations in the
form factors describing the broad $0^-$ component.
The $\eta (1970)$ contribution is close to $90^\circ$ out of phase
with the broad $0^-$, producing a dispersive shaped curve
in Fig. 5(a) due to destructive interference
between the real part  of the $\eta(1970)$ and the imaginary part of
the broad $0^-$ amplitude.

Fig. 5(c) shows the full $0^+$ component, with
clear peaks at 1500 and 1800 MeV.
In Fig. 5(d), the full histogram shows the $2^+$ component, which may
be fitted by a  broad resonance at 2150 MeV, plus a small contribution
from $f_2(1270) \to \rho \rho$.
This panel also displays as the dashed histogram the total incoherent
background, which is included in the fit to data.
In more detail, the $f_2(2150)$ requires almost equal contributions
from decays to $\rho \rho$ and $f_2(1270)\sigma$.
The fitted mass is $2150 \pm 29(stat) \pm 60(syst)$ MeV and the width
$506 \pm 30(stat) \pm 100(syst)$ MeV.
The largest uncertainty in fitting parameters arises from the
large and rapidly varying background of Fig. 5(d).

The mass of the broad $2^+$ component is significantly higher than
the $f_2(1940)$ found in BES I data \cite {BESI} and also higher than
the $f_2(1950)$ of the PDG, which quotes $M = 1944 \pm 12$ MeV,
$\Gamma = 472 \pm 18$ MeV.
This  raises the possibility that the $2^+$ component observed
here consists of two or more unresolved states.
In the production process, the short-range interaction between the
photon and gluons limits the process to production of
$^3P_2$ $\bar qq$ states and excludes $^3F_2$.
Crystal Barrel analysis identifies two $^3P_2$ states, namely
$f_2(1910)$ of the Particle Data Group and $f_2(2240)$ \cite {CBAR},
well identified by polarisation data for $\bar pp \to \pi ^+\pi ^-$,
but incorrectly listed by the Particle Group under the $\bar ss$ state
$f_2(2300)$.
An alternative fit to the $2^+$ signal is made using $f_2(1910)$ and
$f_2(2240)$.
This gives an almost identical fit with log likelihood
better than the first alternative by only 0.8; this is not
statistically significant.
The fit to the $4\pi$ mass projection shows no significant difference
either.
So this is a feature of the data which cannot be resolved
conclusively at the moment.
The Table of results will use the single broad $f_2(2150)$.

\begin{figure}
\begin{center}
\vskip -12mm
\epsfig{file=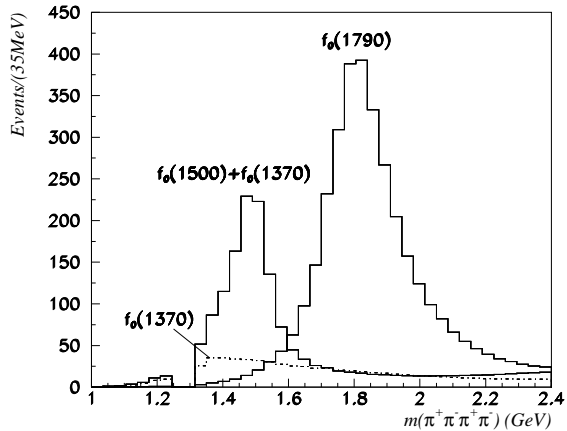,width=7.5cm}
\vskip -6mm
\caption[]{$0^+$ contributions from (i) the coherent sum of
$f_0(1370) + f_0(1500)$, (ii) $f_0(1790)$ and (iii) $f_0(1370)$.
There is a cut from 1245 to 1325 MeV to remove $f_1(1285) \to 4\pi$.
}
\end{center}
\end{figure}

The fit to $0^+$ definitely requires contributions from all of
$f_0(1370)$, $f_0(1500)$ and $f_0(1790)$.
The mass and width of $f_0(1370)$ are fixed at values 1309 and
325 MeV from the recent re-analysis of Ref. \cite {f01370}.
If $f_0(1370)$ is omitted from the fit, log likelihood is worse by
100.3, a highly significant amount.
Because of the rapid increase in $4\pi$ phase space,
its peak in $4\pi$ is at 1390 MeV but it has a long tail in $4\pi$ at
high mass and interferes with all other $0^+$ components.
Consequently, it is not well determined in magnitude.
The possible process $\sigma \to 4\pi$ is inconsistent with the slice
fit of Fig. 3(b) since it would give a broad contribution rising
continuously with mass according to $4\pi$ phase space.

Fig. 6 shows contributions from the coherent sum of $f_0(1370)$ and
$f_0(1500)$.
The effect of $f_0(1370)$ is needed to
broaden the contribution from $f_0(1500)$.
But the resulting tail interferes with $f_0(1790)$.
If the parametrisation of the $f_0(1370)$ component is  varied within
its known errors, the observed mass of the 1790 MeV peak and its full
width at half-maximum remain very steady at $M = 1800 \pm 25$ MeV,
$\Gamma = 230 \pm 30$ MeV.
The observed peak is visually incompatible with
$f_0(1710)$, to which the PDG assigns a width of 137 MeV.
So the observed peak provides independent evidence for the existence of
$f_0(1790)$.
If the mass and width of $f_0(1790)$ are changed to PDG
values for $f_0(1710)$ MeV and the fit is re-optimised, the $4\pi$ mass
projection is very badly fitted, see Fig. 7.

\begin{figure}
\begin{center}
\vskip -10mm
\epsfig{file=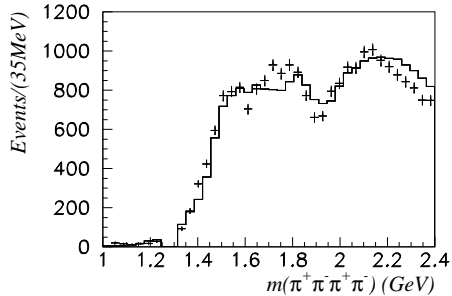,width=6cm}
\vskip -7.5mm
\caption[]{The poor fit to the $4\pi$ mass spectrum if the mass and
width of  $f_0(1790)$ are changed to those of $f_0(1710)$. }
\end{center}
\end{figure}

An important result is that previous evidence for $f_0(2100)$ in
these data has decreased substantially; it is now almost invisible in
the mass projection of Fig. 5(c).
The reason for the change is the new evidence for the
$\eta (1970)$.
The $f_0(2100)$ does appear clearly in E785 data on $\bar pp \to
(\eta \eta )\pi ^0$, and there is independent evidence for it in
Crystal Barrel data \cite {CBAR}.

\begin{figure}
\begin{center}
\vskip -6mm

\epsfig{file=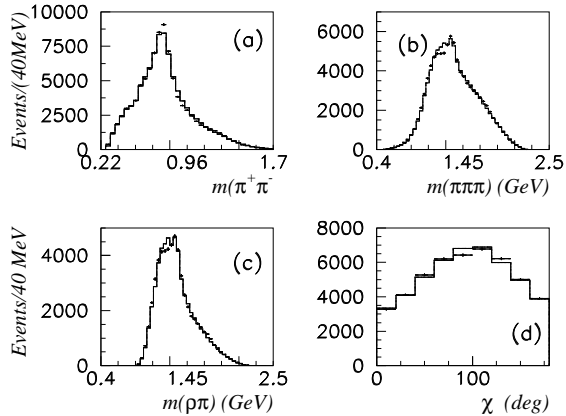,width=7.5cm}\
\vskip -6mm
\caption[]{Fit to mass projection for (a) $2\pi$,
(b) all $3\pi$, (c) $\rho \pi$, with a cut on the $\rho$ from 670 to
870 MeV; (d) the azimuthal angle $\chi$ between the planes of $\rho $
pairs.}
\end{center}
\end{figure}

Fig. 8 shows further details of the fit to data.
Fig. 8(a) shows the fit to the $\rho(770)$ peak.
The $3\pi$ mass projection of Fig. 8(b) and the $\rho \pi$ mass
projection of Fig. 8(c) exhibit the peak at 1320 MeV due
to the $a_2(1320)\rho$ background.
Fig. 8(d) shows the azimuthal angular distribution between
$K_A$ and $K_B$, requiring a large $0^-$ component, which
varies as $\sin ^2 \chi$.
Fig. 8(d) also reveals a substantial flat component.
Unfortunately, the dependence on $\sin ^2 \chi$ alone does not
separate $0^-$, $0^+$ and $2^+$ components cleanly.
Note also that there are two $\rho ^0\rho ^0$ combinations which
interfere coherently; this further hampers a separation of the three
available $J^P$ from $\sin ^2 \chi$ alone.
This is a well known problem in analysis of multi-body final states.
However, the full multi-dimensional correlations between production
angles, decay angles of intermediate isobars and the azimuthal angle
between pairs such as $\rho \rho$ do separate $J^P$ cleanly, as
displayed in Fig. 3.
There is no simple way of displaying the 3-dimensional angular
correlations (or 5-dimensional when two $\rho \rho$ combinations are
included); it is necessary to rely on the detailed amplitude analysis
and the variation of log likelihood with parametrisations of individual
resonances.

\section {More detail on formulae}
The final state $X$ has a radius of interaction between final
$\rho \rho$ or $\sigma \sigma$ configuations of order $R  = 0.7$ fm.
It is necessary to include a form factor describing this
radius of interaction.
The conventional form factor for decays is
\begin {equation}
F_d = \exp (-k^2 R^2/6) = \exp (-\alpha _d k^2),
\end {equation}
where $\alpha _d= (1/6)(R/\hbar c)^2 = (1/6)(R (fm)/0.197321)^2$
with $\alpha _d$ in (GeV/c)$^{-2}$.

For decays with $L=1$, the standard Blatt-Weisskopf centrigual
barrier factor in the amplitude is
\begin {equation}
B = K/[(\hbar c/R')^2 + K^2]^{1/2}.
\end {equation}
Note that the derivation of this formula replaces the centrifugal
barrier by an equivalent square barrier, so $R'$ may be slighly
different from $R$ in the form factor.
However, the Gaussian form factor is only an approximation to unknown
wave functions, so it is convenient to fit data using a single radius
parameter, setting $R = R'$.

There is also a centrifugal barrier and form factor for
the production process, except that $R$ takes a much smaller
value; the parameter $\alpha _d$ of Eq. (1) is replaced by a much
smaller parameter $\alpha _p$.
The intensity of the $J^P = 0^-$ signal as a function of mass is
given by the square of the form factor and centrifugual barrier
factor, multiplied by a factor $P$ for phase space in the production.
The result is the factor $P^3$ quoted above, multiplied by a weak
form factor $F_p = \exp (-\alpha _p P^2)$.

A general remark concerns the mass of the $\rho \rho$ peak.
If the width of the $\rho$ were zero, the P-wave intensity
for $\rho \rho$ would rise from threshold as $k^3$ where $k$ is the
momentum of each particle in the $\rho \rho$ system.
A centrifugal barrier and decay form factor of conventional radius
$\sim 0.7$ fm would make the $\rho \rho$ and $\omega \omega$
intensities both peak in the range 1700-1800 MeV.
The width of the $\rho$ spreads this peak out further than that
for $\omega \omega$.
It is the factor $P^3$ for the production cross section from a
point-like source which reduces the mass of the peak to $\sim 1600$ MeV.

It is necessary to fold the width of the two $\rho$ into the evaluation
of $4\pi$ phase space. This is given by
\begin {equation} \rho _{4\pi}(s) =
\int ^{(\sqrt {s} - 2m_\pi )^2} _{4m^2_\pi}
\frac {ds_1}{\pi} \int ^{(\sqrt {s} - \sqrt {s_1})^2}_{4m^2_\pi}
\frac {ds_2}{\pi}
\frac {8|k|\,|k_1| \, |k_2|} {\sqrt {s \, s_1 \, s_2}}
|T_1(s_1)|^2 |T_2(s_2)|^2 \exp (-2\alpha _d k^2),
\end {equation}
where $k_1$ and $k_2$ are momenta of pions from the decay of each
$\rho$ in its rest frame, and $k$ stands for the momenta of the $\rho$
in the rest frame of the resonance $X$.
The amplitude $T(s)$  is the usual Breit-Wigner amplitude for the
$\rho$.
These integrals are done numerically and then parametrised by smooth
functions of $s$.

\subsection {The effect of thresholds on resonances}
It is common practice to fit resonances with Breit-Wigner amplitudes
with constant width.
However, the Flatt\' e formula shows this to be only an approximation.
That formula describes the amplitude by
\begin {equation}
T=F(s)/[M^2 - s - i \sum _i g^2_i \rho _i(s)],
\end {equation}
where $F(s)=F_pF_d$, $g_i$ is the coupling constant and $\rho _i$ the
Lorentz invariant phase space for each channel $i$.

Eqn. (4) however requires a further term if the amplitude is to be
described correctly by an analytic function of $s$.
It has been known since the 1950's that the full form for $T$ is
\begin {eqnarray}
T &=& F(s)/[M^2 - s -m(s) - i\sum g^2_i\rho _i(s) ]
\\ m(s) &=& \frac {(s - M^2)}{\pi}P \int \frac {\sum _i g^2_i \rho
_i(s')\, ds'}{(s' - s)(s' - M^2)}.
\end {eqnarray}
The additional term $m(s)$ describes the $s$-dependence of the real part
of the amplitude going with the $s$-dependence of the imaginary part.
It is essential in fitting the present data near the $\rho \rho$
threshold; without it, the fit is very bad, see Fig. 11(b) below.
The mass dependence of $m(s)$ is shown in Fig. 9.
The new parametrisation of the broad $0^-$ amplitude supercedes
primitive alternatives used in Refs. [4] and [5].

\begin{figure}
\begin{center}
\vskip -1.5cm
\epsfig{file=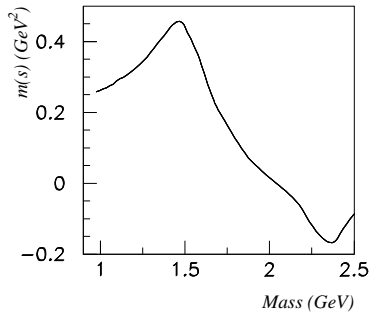,width=6cm}\
\vskip -0.6cm
\caption[]{The dispersive term $m(s)$ for the broad $0^-$ contribution.}
\end{center}
\end{figure}

A general comment is needed on the role of $M^2$ in Eqs. (4)--(6).
There may be further decay channels, presently unknown.
Also the exponential form factor of Eq. (1) is a conventional educated
guess; it is needed to make the dispersion integral of Eq. (6)
converge.
Realistically there are uncertainties in the high energy behaviour of
this integral, well known as a renormalisation effect.
The value of $M^2$ may be viewed as a way of applying an empirical
correction to $m(s)$ for such effects.
A single broad resonance has been used in the earlier work by Bugg,
Dong and Zou \cite {Broad0M}.
In the earlier work $M$ was 2.19 GeV; now it moves to 2.04 GeV.

Fig. 10 shows that similar threshold behaviour is observed in $\eta
\pi\pi$ , $K\bar K^*$, $\omega \omega$ (from Mark III) \cite
{Wermes}, $K^*\bar K^*$ and $\phi \phi$.
Fig. 10(a) shows the broad $0^-$ component fitted to
$\gamma (\eta \pi \pi)$ data, the sum of $\eta \sigma$ and $a_0\pi$.
Fig. 10(b) shows the coherent sum of $\eta
(1440)$ and the broad $0^-$ fitted to the $K\bar K^*$ channel in an
accompanying paper on  $J/\Psi \to \gamma (\eta \pi ^+\pi ^-)$ and
$\gamma (K^\pm K^0_S \pi ^\mp )$ \cite {iota}.
Fig. 10(c) shows what is observed with $J^P = 0^-$ in $\gamma \rho
\rho$.
A further threshold peak (not shown) is observed in BES II data
for $J^P = 0^+$ on $J/\Psi \to \gamma \omega \phi$ \cite {omphi}.
The natural explanation of all these processes is that
$J/\Psi \to \gamma GG$, where $G$ are gluons.
These produce pairs of vector mesons via colour
neutralisation.
In turn, these may de-excite to lower mass
configurations, e.g. $[\rho \rho]_{L=1} \to [\eta \sigma ]_{L=1}$.
This can explain why the $\eta \sigma$ intensity peaks strongly at the
same mass as $\rho \rho$.
The inference is that all these peaks may be threshold cusps.
Such cusps are well known in other processes.
Examples are threshold peaks in $\pi d \to NN$ and
$\bar pp \to \bar \Lambda \Lambda$ \cite {cusps}.
The other possibility is that one or more of these peaks is resonant.
\begin{figure}
\begin{center}
\vskip -1.5cm
\epsfig{file=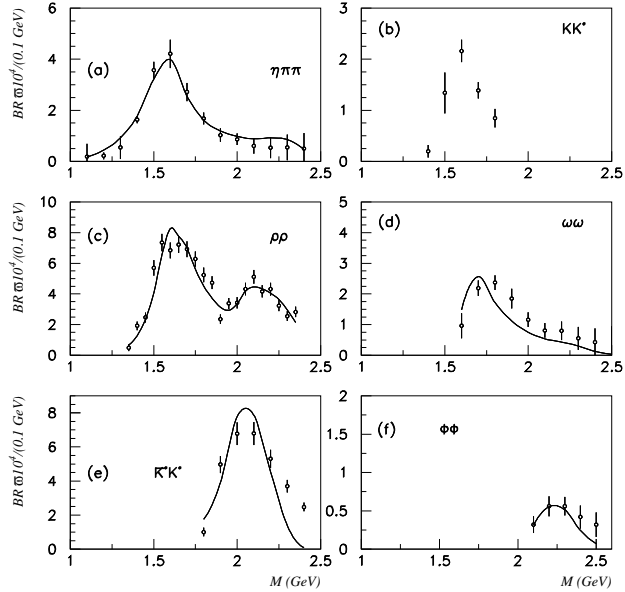,width=9cm}\
\vskip -6mm
\caption[]{Fit to mass projections for $0^-$ signals in 6 channels.}
\end{center}
\end{figure}

The data have been fitted using Eqs. (5) and (6) including the effects
of cusps in all coupled channels.
The coupling constants $g^2_i$ are adjusted iteratively to reproduce
branching fractions of every channel in Fig. 10.
As a result, the form of the broad $0^-$ is tightly constrained.
The dispersive term associated with the opening of the $\rho \rho$
threshold plays a crucial role.
Without it, the fit to data, shown on Fig. 11(b), is hopelessly bad.
Fig. 11(a) shows the Argand diagram for Eq. (5).
Note that what is plotted is the amplitude without the
form factor $F(s)$ of the numerator.
The reason for this is that the factors $P$ for production and $k$
for decay are built into the tensor expressions for amplitudes,
as well as the associated form factors.
The intensity observed in data at low momentum in Fig. 10(b) is
inflated by the factor $P^3$ for production.
Note too that there are two interfering $\rho \rho$ combinations in
Fig. 11(b).

Approaching 1900 MeV, the magnitude of the amplitude on Fig. 11(a)
is decreased by the opening of the strong $K^*\bar K^*$ threshold.
As this channel decreases above 2200 MeV, the amplitude recovers.
Note that this threshold does  {\it not} account for the structure
near 1900 MeV which has been fitted by the additional $\eta (1970)$.

On the Argand diagram of Fig. 11(a), there is not a complete loop
corresponding to a resonance in the mass range near 1600 MeV.
It is being fitted as a simple threshold cusp.
Below the threshold, the real part of the amplitude moves positive.
Over the threshold, the imaginary part of the amplitude increases
rapidly.
Above the threshold, the real part of the amplitude decreases
slowly.
The result is a half loop, characteristic of a cusp.
\begin{figure}
\begin{center}
\vskip -2cm
\epsfig{file=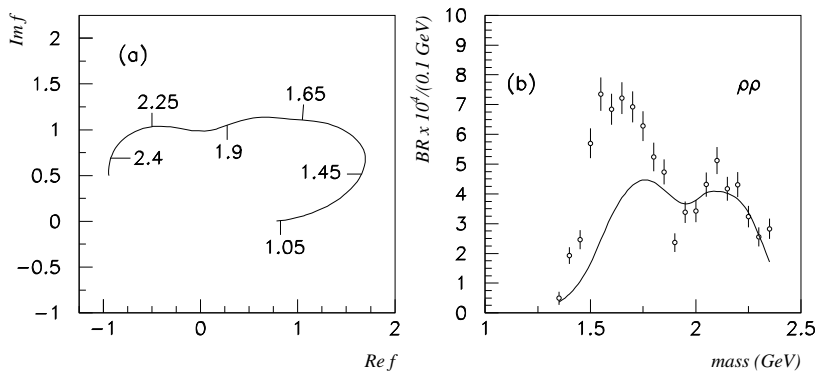,width=12cm}\
\vskip -6mm
\caption{(a) Argand diagram for the broad $0^-$; masses are shown in
GeV;
(b) the bad fit to the $0^-$ mass projection with the dispersive term
$m(s)$ omitted.}
\end{center}
\end{figure}

However, from Eqs. (5) and (6) and fitted parameters, it turns out
that there is a broad pole at $M=M_0 -i\Gamma /2 = 1.564 \pm
0.012 - i(0.139 -i 0.035)$ GeV; errors are systematic and are
estimated from the parametrisation of the $\rho \rho$ channel and its
extrapolation into the complex $s$-plane.
There is no discernable pole due to the weaker P-wave $\omega \omega$
threshold which peaks at 1750 MeV.
There are more distant poles, including one at $1.862 - i0.210$ GeV
associated with the opening of the $K^*\bar K^*$ channel.

For present data, the parameter $\alpha _d$ controlling the form factor
in Eq. (1) optimises at $2.25 \pm 0.25$ GeV$^{-2}$, corresponding to
a reasonable radius of interaction for decays, $0.73 \pm 0.04$ fm.
The value of $\alpha _p$ for the production process optimises at
0.06 GeV$^{-2}$, corresponding to a radius of interaction 0.12 fm.
As an aid to future work where the broad $0^-$ is relevant,
the fully annotated Fortran code used for it is available directly
from the author.
This code includes numerical parametrisations of $\rho _i(s)$ for
$\alpha _d$ in steps of 0.25 GeV$^{-2}$.
For values of $\alpha _d$ or $\alpha_p$ different to those used here, it
will be necessary to re-optimise $g^2_i$ fitted to all relevant sets of
data.

\section {Branching fractions}
Table 1 summarises the significance levels for each component,
measured by changes $\Delta S$ in log likelihood when the
component is removed from the fit and all other components are
re-optimised.
Statistically, $\chi ^2$ is approximarely twice $\Delta S$.
However, experience shows that signals less than 20 in log likelihood
are of questionable significance because of the possibility of
systematic errors.
Contributions in this category are the weak decays of $f_0(1790)$ and
$f_0(2100)$ to $\rho \rho$ and also $f_2(1565)$, which produces
a change of log likelihood of just 11.9.

\begin{table}[htb]
\begin {center}
\begin{tabular}{ccc}
\hline
channel & $\delta (ln L)$ & Branching fraction $\times 10^4$ \\\hline
$\eta (2190) \to \rho \rho$ & $> 1000$ &  \\
$\eta (1970) \to \rho \rho$ & $233.5$ & $2.0 \pm 0.3$ \\
Both $0^-$                   & &$13.7 \pm 1.4$            \\
$f_0(1370) \to \sigma \sigma$ & $100.3$ & $1.1 \pm 0.4$   \\
$f_0(1500) \to \sigma \sigma$ & $163.8$ & $1.5 \pm 0.4$  \\
$f_0(1790) \to \sigma \sigma$ & $140.0$ & $5.8 \pm 1.2$ \\
$f_0(1790) \to \rho \rho $ & $3.0$ &  \\
$f_0(2105) \to \sigma \sigma$ & $25.2$ & \\
$f_0(2105) \to \rho  \rho$ & $6.0$ & \\
$f_2(1565) \to \rho  \rho$ & $11.9$ & $0.7 \pm 0.3$ \\
$f_2(2150) \to \rho  \rho$ & $70.2$ &  \\
$f_2(2150) \to f_2(1270)\sigma $ & $178.5$ &  \\
$f_2(2150) \to \rho \rho + f_2(1270)\sigma $ & $190.5$ & $3.1 \pm 0.5$
\\\hline
\end{tabular}
\caption{Changes in log likelihood when each
channel is dropped from the fit and remaining contributions are
re-optimised.
Column 3 shows branching fractions for $J/\Psi \to \gamma X$, $X \to
\pi ^+\pi ^-\pi ^+\pi ^-$.
Missing elements are not quoted because of large uncertainties due to
interferences.
Errors include systematic errors on the overall normalisation.}
\end {center}
\end{table}

The total number of selected events is 23591.
The total background fitted up to a mass of 2.4 GeV is 27.3\% of all
events, leaving a total signal of 72.7\%.
The corresponding branching fraction for all events is
\begin {equation}
J/\Psi \to \gamma (\pi ^+\pi ^-\pi ^+\pi ^-) = (25.5 \pm 2.6(syst))
\times 10^{-4}, ~M(4\pi) < 2.4~ {\rm GeV}.
\end {equation}
The error is systematic and arises mostly from uncertainty in the
experimental background.
In determining branching fractions given in Table 1,
interferences within one channel between the two possible $\pi ^+\pi
^-$ combinations are included, but interferences between different
channels are dropped.
This is because interferences within one resonance should be universal,
but interferences with other channels vary from process to process.
Note that the PDG quotes branching fractions for all $\rho \rho$ and
$\sigma \sigma$ charge states.
These are conventionally obtained by multiplying the branching
fractions of Table 1 by a factor 3 for $\rho \rho$ and a factor $9/4$
for $\sigma \sigma$.
However, it is likely that interferences between different
charge configurations of a single resonance will be different between
$\rho ^0\rho ^0$ and $\rho ^+\rho ^-$.
Chen et al. \cite {Chen} have estimated the magnitudes of such effects.

The fitting program itemises the magnitudes of all individual resonances
and all interferences between them.
A warning is that interferences are quite large within
one $J^P$.
There is destructive interference between the two decay modes
of $f_2(2150)$, so the branching fraction is quoted only for their
coherent sum.
There is also destructive interference between the broad $0^-$
amplitude and $\eta (1970)$, so the table quotes the total $0^-$
branching ratio.
Within the $0^+$ sector, the branching fractions of $f_0(1500)$
and $f_0(1770)$ are quite well determined, but there is a poorly
determined interference with the tail of $f_0(1370)$.
The magnitude of any possible $f_0(2100)$ contribution is difficult
to determine because of interferences with the high  mass tails of
all of $f_0(1370)$, $f_0(1500)$ and $f_0(1790)$, and is therefore
not quoted.

There is no direct interference between $J^P = 0^-$ and $0^+$.
Interferences between $0^+$ and $2^+$ are also quite small.
The summed events for $0^-$, $0^+$ and $2^+$ add up to 17166
events, to be compared with the total of 17146 events after
background subtraction.

\begin{figure} [htb]
\begin{center}
\vskip -1cm
\epsfig{file=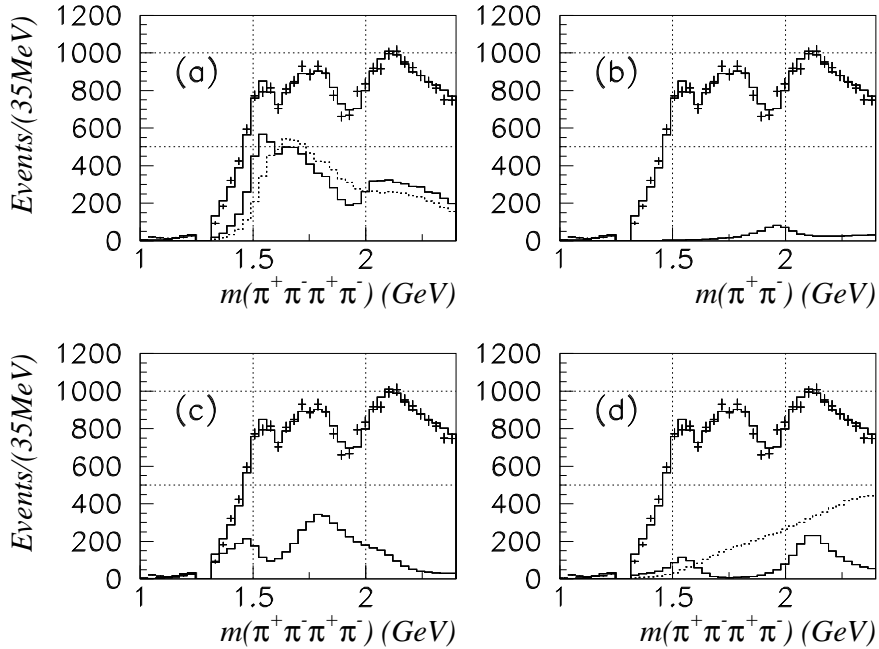,width=12cm}\
\vskip -6mm
\caption{The fit, including an $\eta (1560)$, to the $4\pi$ mass
projection and magnitudes of fitted components in the same format as
Fig. 5.}
\end{center}
\end{figure}

\section {Subtle evidence for a $0^{-+}$ resonance near
1560 MeV}
The $4\pi$ mass projection shown on Fig. 5 is not fitted perfectly
from 1500 to 1650 MeV.
There is one low point at 1610 MeV.
Moving the edges of bins or changing the bin width does not
alter this point significantly.
There is also some evidence for a similar dip in DM2 data
\cite {DM2} and a dip in Mark III data at 1650 MeV \cite {Scott}.

It is unreasonable to invent a new narrow resonance to fit the
single low point at 1610 MeV.
However, the observed pole in the broad $0^-$ component is already
a hint that the $\rho \rho$ threshold peak is resonant.
This is what would be called in today's terminology a dynamically
driven $\rho \rho$ resonance.
It is quite likely to mix with the $n\bar n$ radial excitation
expected in this general mass range, for example the radial excitation
of $\eta (1295)$ if it exists.
Suppose there is such a $0^-$ resonance in the range
1550--1700 MeV.
Experience with other sets of data, for example Crystal
Barrel data, Refs. \cite {f01370} and \cite {wrho}, is
that a resonance at or close to a broad threshold is in practice
described well by a simple pole with a high mass dispersive tail
given by the term $m(s)$ of Eq. (6).
That is, a good approximation can be obtained by adding a simple
Breit-Wigner pole to the dispersive amplitude used so far.

It is instructive to fit the data this way, scanning the mass and
width of the additional pole.
It turns out that this gives a distinct improvement in the fit to
the $4\pi$ mass projection and the dip at 1610 MeV.
It gives an improvement in $2 \times$ log  likelihood (which
follows a distribution very close to $\chi ^2$) of 194 for four extra
fitting parameters, statistically a 10.6 standard deviation effect.
Systematic errors quoted for mass and width cover all variations
as masses and widths of $f_0(1500)$ and $f_0(1790)$ are varied over
their allowed ranges from earlier data; systematic errors
also cover uncertainty in the dispersive contribution of the $\eta
\pi\pi$ channel.

The improvement in the fit comes partially from the fit to the $4\pi$
mass spectrum.
However, it also comes from systematic improvements amongst decays to
$0^+$, $2^+$ and $0^-$ partial waves.
The fitting program takes the trace over both allowed spin states
of the photon.
There are interferences between $0^-$ and $2^+$
and between $0^+$ and $2^+$, dependent in subtle ways on
three-dimensional angular correlations, hence giving information
on relative phases.

The new fit is shown in Fig. 12 in precisely the same format as
Fig. 5.
There is an obvious improvement in the fit to the $4\pi$ mass
spectrum from 1500 to 1650 MeV, particularly near the dip.
The signal due to $f_0(1500)$ in Fig. 12(c) decreases slightly
and the dip before $f_0(1790)$ is also slightly lower.
No change is required to the parameters of $f_0(1790)$.
The mass of $\eta (1970)$ does not change, but its width
decreases from $210 \pm 25(stat) \pm 60(syst)$ to
$165 \pm 25(stat) \pm 60(syst)$ MeV.
A knock-on effect of this reduction is that the width of
the broad $2^+$ component at 2150 MeV decreases from 506 MeV to
$300 \pm 75$ MeV.
The evidence for $f_0(2100)$ increases slightly on Fig. 12(c).
A definite $f_2(1565)$ appears in Fig. 12(d).
Except for this state, branching fractions remain within the errors
shown in Table 1.
The branching fraction for $f_2(1565)$ rises to $(1.3 \pm 0.3) \times
10^{-4}$, nearly double that of Table 1.

\begin{figure} [htb]
\begin{center}
\vskip -1.5cm
\epsfig{file=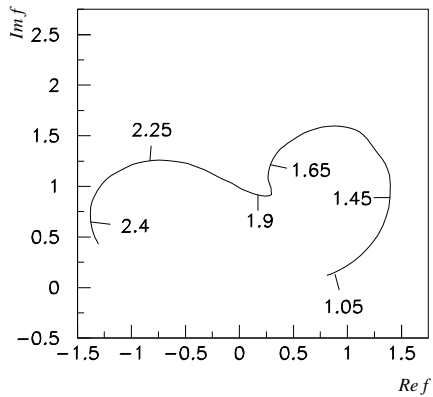,width=7cm}\
\vskip -6mm
\caption{The Argand diagram for the coherent sum of the 1560 MeV
$0^-$ pole and the dispersive cusp; masses are shown in GeV}
\end{center}
\end{figure}

The Argand diagram for the coherent sum of the 1560 MeV pole and
the dispersive cusp is instructive and is shown in Fig.
13.
The pole term is $\sim 35^\circ$ out of phase with the
dispersive term, and the result is to increase the speed
with which the amplitude describes the Argand loop of
Fig. 13.
As one sees from Fig. 12(a), the direct effect on the whole $0^-$
amplitude is  to produce half the required dip at 1610 MeV.
The remaining dip is created by a decrease in the $0^+$
amplitude between $f_0(1500)$ and $f_0(1790)$.
The increased $f_2(1565)$ intensity in Fig. 12(d) and its
interference with the $0^+$ amplitude contributes to the fit to the
$4\pi$ mass spectrum near 1560 MeV.

The coherent sum of the additional pole and the broad $0^-$
has been parametrised as a function of $s$ and extrapolated into the
complex plane.
A pole is observed at $M = 1560 \pm 12 - i(143\pm 35) $ MeV,
close to that of the broad $0^-$ alone.

The improvement in log likelihood when the $0^-$ pole at 1560 MeV
is included is larger than for most of the other contributions in
Table 1, so the evidence for this resonance must be taken seriously.
In summary, there is no doubt about the existence of a strong
$\rho \rho$ peak at $\sim 1600$ MeV.
However, evidence for resonant behaviour is hidden in the angular
correlations.
It would be valuable to have further $4\pi$ data in which the $0^-$
component interferes with another well defined component;
central production in the Compass experiment is one such possibility.

\section {A conflict with $\gamma \omega \omega$ analysis}
A BES II paper on $J/\Psi \to \gamma \omega \omega$ claims a large and
almost pure $\eta (1760)$ signal in $\omega \omega$ \cite {gww}.
No such resonance is observed in the present data.
This raises a problem.
There is a well known relation, coming from SU(2) symmetry,
that an $I=0$ resonance should have equal couplings to
$\omega \omega$ and $\rho ^0\rho ^0$.
Physically, light quarks do not discriminate between charges and
therefore couple equally strongly to $\omega \omega$ and
$\rho ^0\rho ^0$.
Because there are three charge states for $\rho \rho$, the
relation is normally written $g^2(\rho \rho ) = 3g^2(\omega\omega)$.
This relation applies equally to $q\bar q$ states, hybrids and
glueballs, since it springs from SU(2) symmetry, which is obeyed by all
these configurations.
The Appendix gives an elementary derivation of
this result for a $q\bar q$ state.
Similar algebra generalises the result to all three types of state.

The branching fraction quoted for production of $\eta (1760)$ in
the $\gamma \omega \omega$ data is
$(1.98 \pm 0.08(stat) \pm 0.32 (syst)) \times 10^{-3}$.
This is larger than the branching fraction in present data for
production of $\rho ^0\rho^0$ with $J^P = 0^-$ over the
{\it entire} mass range: $(1.37 \pm 0.14) \times 10^{-3}$.
It should lead to a huge peak with resonant phase variation
interfering with the broad $0^-$ component.
The present data are in complete disagreement with that
possibility.
Furthermore, such a large resonance is also in conflict with all
earlier data from BES 1 \cite {BESI}, Mark III \cite {MarkIII} and
DM2 \cite {DM2}.
The DM2 group did claim a small $\eta (1760)$ signal originating from
the peak near this mass in the total $4\pi$ mass projection; however,
this had a branching fraction an order of magnitude smaller than
BES II claim.
Furthermore, DM2 did not try a $0^+ \to \sigma \sigma$ contribution
which fits the peak in present work.

The likely explanation of the conflicts is that the $0^-$ signal in
$\omega \omega$ is not a resonance, but a threshold cusp.
Note that there is no phase information identifying resonant
behaviour of $\eta (1760)$ in BES data.
It was simply assumed that the peak is resonant.
A threshold cusp  arises because the $\omega \omega$ intensity
rises initially from threshold as $k^3$, but peaks near 1700 MeV due to
the form factor $\exp -(\alpha _d p^2)$; here $k$ is the momentum of
each $\omega$ in the $\omega \omega$ rest frame.
The precise position of the peak depends rather strongly on
$\alpha _d$.
The BES publication does not give details of the form factor assumed
for the P-wave.

This explanation may be tested by refitting the
$\omega \omega$ mass projection of BES II data using the $\omega \omega$
decay of the broad $0^-$ fitted to present data.
The $\eta (1970)$ component should also obey the SU(2) relation between
$g^2_{\omega \omega }$ and $g^2_{\rho ^0 \rho ^0}$.
This constraint is applied by fixing relative magnitudes of $\eta
(1970)$ and the broad $0^-$ to the value fitted to present
data.
However, the phase of $\eta (1970)$ relative to the broad $0^-$
is fitted freely, since it is determined by multiple
scattering amongst all components, which are different in $\omega
\omega$ and $4\pi$ final states.

Decays of $f_0(1790)$ to $\omega \omega$ are also included. but are
limited in intensity to $20\%$ of the broad $0^-$.
This is the maximum which can be fitted to $f_0(1790) \to \rho ^0 \rho
^0$ in present data.
Decays of the $0^+$ state produce a $\cos ^2 \chi$ decay distribution
(where $\chi$ is the angle between $\omega $ decay planes)
which combines with an equal amount of $0^-$ to produce a flat
component in the $\chi$ distribution.
This allows the $0^-$ component of BES II data to rise to $\sim 46\%$
of those data.

There may be small $2^+$ contributions from $f_2(1565)$ and
$f_2(1910)$/$f_2(2150)$.
Figs. 5(d) and 12(d) show that these are almost zero at 1750 MeV.
They can only be determined from the $\gamma \omega \omega$ data
using the full angular correlations.
They are therefore omitted from the test made here.

\begin{figure}
\begin{center}
\vskip -2cm
\epsfig{file=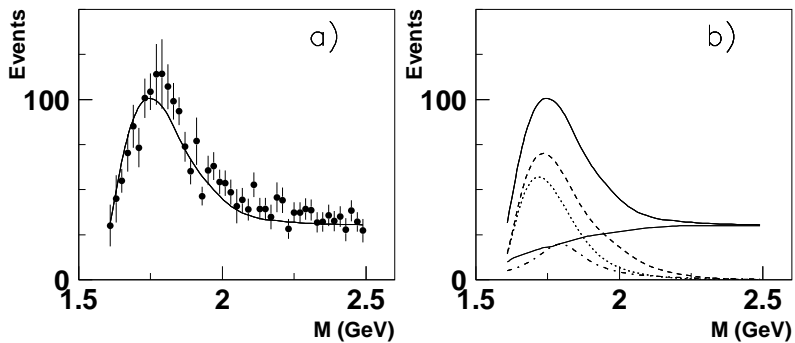,width=12cm}\
\vskip -6mm
\caption{The fit to BES II $\gamma \omega \omega $ data as a threshold
cusp.
In (b), the upper full curve shows the total contribution;
the dotted curve shows the broad $0^-$ alone, and the
dashed curve its coherent sum with $\eta (1970)$.
The chain curve shows the $f_0(1790)$ intensity and the lower
full curve the experimental background scaled from the BES II
publication.}
\end{center}
\end{figure}

Fig. 14(a) shows the resulting fit to the $\omega \omega$ mass
spectrum.
The optimum fit requires some reduction of the parameter $\alpha _d$
of the form factor for decays to 2.0 GeV$^{-2}$.
The fit is close to the data, showing that they may indeed be fitted
by a non-resonant threshold cusp.
The optimum fit would require $\alpha _d < 2.0$ GeV$^{-2}$, but that
would create an inconsistency with the $\rho \rho$ data.
Details of the components are shown on Fig. 14(b).
The full curve is the same as in Fig. 14(a) as a reference.
The dotted curve shows the contribution from the broad $0^-$,
and the dashed curve shows its coherent sum with the $\eta (1970)$.
The overall effect of $\eta (1970)$ is quite small.
The chain curve shows the $f_0(1790)$ contribution, fitted with the
required phase space for S-wave decays.
The broad $0^-$ makes a small but significant
contribution above 2 GeV; BES II fitted that mass range purely as
background.
To allow for this, the $s$-dependence of the background is taken from
the BES publication, but its absolute magnitude is fitted freely and
is shown by the lower full curve on Fig. 14(b).

The fit in Fig. 14 is made without considering the absolute magnitude
of the branching fraction for $J/\Psi \to \gamma \omega \omega$.
The BES II publication quotes branching fractions in their
Table 2 adding to $(2.77 \pm 0.12) \times 10^{-3}$ of all $J/\Psi$
decays.
It is a matter of conern that this is considerably larger than
quoted by DM2 \cite {DM2}, $1.42 \pm 0.2 \pm 0.42$ and Mark III, $1.76
\pm 0.09 \pm 0.45$ \cite {MarkIII}.
Also, one would expect the branching fraction for $\rho \rho$
after integration over all masses to be roughly the same as for
$\omega \omega$; the $\rho \rho$ peak is simply spread out compared
with $\omega \omega$ by the width of the $\rho$.
The disagreements over branching fractions raises the question of
backgrounds in the BES data.
Those used in the BES publication were estimated by Monte Carlo
simulations, which really need checking against the standard sideband
subtraction procedure.
Also the BES II analysis chose the best of 15 combinations of photons
fitting $\gamma \omega \omega$; wrong combinations will make a
significant background, but this background is not discussed in the
BES publication.

\section {Summary and Conclusions}
In summary, the new data confirm earlier detection of $0^+$ states at
1500 and  1790 MeV in this channel.
There is a broad $2^+$ signal at $\sim 2150$ MeV decaying to both $\rho
\rho$ and $f_2(1270)\sigma$, but this may be an unresolved combination
of more than one state.

For $J^P= 0^-$, the new data give a distinctly
improved definition of the broad $0^-$ component.
The exponential form factor for decays is  well determined, with
$\alpha _d = 2.25 \pm 0.25$ (GeV/c)$^2$, corresponding to a radius of
interaction of $0.73 \pm 0.04$ fm for decays.
The exponential form factor for production is also well determined,
with very small $\alpha _p = 0.06$, corresponding to a radius of
interaction of $0.12$ fm for the production process; this
is consistent with the expected point-like interaction.
There is also a definite dispersive shaped effect with a
resonant phase variation with respect to the broad $0^-$
component, requiring the presence of a $0^-$ resonance with mass $1970
\pm 25(stat) \pm 60(syst)$ MeV and width $210 \pm 25(stat) \pm
60(syst)$ MeV, in fair agreement with a less definitive Crystal Barrel
observation \cite {CBAR}.

There is subtle but persuasive evidence that the observed
$\rho \rho$ threshold peak with $J^P = 0^-$ resonates at
$\sim 1560$ MeV.
Because there is only limited phase information from interference with
$f_2(1565)$ (which in turn interferes with $f_0(1500)$ and
$f_0(1790)$), confirmation of a resonance is needed in other data
where some well defined component plays the role of an interferometer.

In both Crystal Barrel data on (a) $\bar pp \to \eta \pi ^0 \pi ^0$
\cite {zeromA} and (b) $\bar pp \to 3\eta$ \cite {zeromB}, there is
evidence for a $J^{PC} = 0^{-+}$ resonance in $f_0(1500)\eta$ at
$2285 \pm 20$ MeV in (a) and at $2320 \pm 15$ MeV in (b).
The latter is a particularly simple channel exhibiting a clear peak.
One can now identify a Regge trajectory of conventional slope for
$\eta (550)$, $\eta (1295)$, $\eta (1970)$ and $\eta (2320)$.
The possible $\eta (1560)$ falls 100 MeV below this trajectory.
However, its mass is subject to a systematic error of $\pm 60$ MeV at
present.

\vspace{0.5cm} I wish to acknowledge financial support from the Royal
Society and Queen Mary College. I am also grateful for help from
members of the BES collaboration in processing data and running the
Monte Carlo simulation of acceptance and backgrounds.

\section {Appendix}
The $I=0$ $\bar nn$ combination is
$(u\bar u + d\bar d)/\sqrt {2}$.
Its decay is via the production of a $^3P_0$ pair
$(u\bar u + d\bar d)/\sqrt {2}$.
The final states is $(u\bar u + d\bar d)(u\bar u + d\bar d)/2$.
An $\omega \omega $ pair has this compositon by inspection.

What about $\rho \rho$ decays?
The isospin Clebsch-Gordan coefficients for an $I=0$ resonance
coupling to two $I=1$ particles is
$(\rho ^+\rho ^- + \rho ^- \rho ^+ - \rho ^0 \rho ^0)/\sqrt {3}$,
where $\rho ^+ = u\bar d$, $\rho ^0 = (u\bar u - d\bar d)/\sqrt {2}$
and $\rho^- = -\bar u d$. So
\begin {eqnarray}
(\rho ^+\rho ^- - \rho ^0 \rho ^0 +\rho ^-\rho ^+)/\sqrt {3} &=&
-u\bar d \bar u d - (u\bar u - d\bar d)(u\bar u - d\bar d)/2
- \bar ud u\bar d \\
&=&(-2u\bar d\bar u d - u\bar u u\bar u - d\bar d d\bar d +
u\bar d \bar u d + \bar u d u\bar d - 2\bar u d u \bar d)/2  \\
&=&-(u\bar u u\bar u + d\bar d d\bar d + u\bar d \bar u d
+ \bar u d u \bar d)/2 \\
&=& -(u\bar u + d\bar d)(d\bar u + d\bar d)/2.
\end {eqnarray}
So the decay of $I=0$ $\bar nn$ produces the combination
$(\rho ^- \rho ^0 -\rho ^+\rho ^- - \rho ^-\rho ^+)$,
i.e. the same $\rho \rho ^0$ intensity as $\omega \omega$
and also $\rho ^-\rho ^+$.

\par
{\hskip 0.4cm}

\end{document}